\documentclass[namedreferences]{kluwer}
\usepackage{epsfig}

\begin{document}

\begin{article}

\begin{opening}

\title{A Photometric System for Metallicity Mapping in Galaxies}

\author{V. \surname{Vansevi\v {c}ius} \email{wladas@itpa.lt}}
\institute{Institute of Physics, Go\v {s}tauto 12, Vilnius 2600, Lithuania \\
National Astronomical Observatory, Osawa 2-21-1, Mitaka, Tokyo 181, Japan}
\author{A. \surname{Vazdekis} \email{vazdekis@mtk.ioa.s.u-tokyo.ac.jp}}
\institute{Institute of Astronomy, University of Tokyo, Osawa 2-21-1, Mitaka, Tokyo 181, Japan}
\author{F. \surname{Prada} \email{fprada@bufadora.astrosen.unam.mx}}
\institute{Instituto de Astronom\'{\i}a, Ensenada, Observatorio Astron\'omico
Nacional, U.N.A.M., Mexico}

\begin{abstract}

A new photometric system suitable for deep, precise and quick metallicity
mapping in galaxies is proposed. We find a linear correlation between our
metallicity index and the $Mg_{2}$ index for stellar, globular cluster,
and early-type galaxy spectra, and model spectral energy distributions
of the simple stellar populations.

\end{abstract}


\end{opening}

\section {Introduction}

To understand how galaxies are formed and evolve we need to investigate
their stellar populations and to derive their main parameters
such as metallicity and age. During the last decade determination of the
metallicity has been performed using spectroscopic approach in
particular, by measuring the prominent absorption feature around $518~nm$,
the so called $Mg_{2}$ index (Burstein et al. 1984). It was shown that
$Mg_{2}$ is a good metallicity indicator for globular clusters. This feature
has been also measured extensively by many authors for ellipticals
(see Trager et al. 1998 and references therein). Different theoretical
approaches also show that this feature is a function of metallicity
(e.g., Mould 1978; Barbuy 1994; Casuso et al. 1996).
However spectroscopic observations are strongly limited by the large telescope
time consumption because of a high signal to noise ratio ($S/N$) requirement.
This is why most of the studies deal with the innermost regions of the galaxies
and only a few attempts to reach the galaxy effective radius were made
(e.g. Saglia et al. 1993). Another limitation arises from the fact that the
observations are performed via long-slit spectroscopy, thus lacking a possibility
to {\em map} the galaxies. Such maps are crucial for full understanding of the
spatial distribution of metallicity within a galaxy and thus for constraining the
galaxy formation models. Spectroscopic mapping was attempted by Peletier et al. (1999)
using Integral Field Spectroscopy. However this study is restricted to the very
innermost regions of bright nearby galaxies. An alternative way is to use three
well tuned narrow-band filters for $Mg$ feature and continuum on the both sides of it
(Beauchamp \& Hardy 1997). However, this system is rather telescope time consuming.
The main purpose of this work is to define and investigate a new photometric system
suitable for the efficient mapping of the $Mg~(518~nm)$ feature.

\section {Definition of the Photometric System}

A photometric system suitable for an accurate and quick mapping of the
metallicity in galaxies should satisfy the following requirements:
possess linear relation of a photometric index with the well defined
spectroscopic metallicity indicator; allows wide coverage in radial velocity
($v_{rad}$) with a single set of filters; maintains high stability of the response
and precision of measurement; is robust against the influence of the galaxy's
internal radial velocity; retains homogeneity of the parameters over the wide field.
Taking into account these requirements we have designed a new photometric
system which is composed of two filters: narrow-band
$\left(W1,\delta \lambda =8~nm\right)$ and wide-band
$\left( W2,\delta \lambda =45~nm\right)$, both centered at $519~nm$.
The present configuration is optimized for the observations with $v_{rad}$
ranging from $\sim-500$ to $\sim2000~km/s$. However, this system could be
extended to higher $v_{rad}$ if a new narrow-band filter with redshifted
central wavelength is added. In the present investigation we limit ourselves
with a theoretical set of filters which, for the sake of simplicity, have
rectangular transmission function. An application of the transmission curves
of real filters would not affect the main conclusions of this work, but
generality of the discussion would be lost.

The color index $W$ is defined as follows:
\begin{equation}
W=\left( W1-W2\right) +C,
\end{equation}

\noindent
where $W1,W2$ and $C$ are defined as:

\begin{equation}
W1=-2.5\lg \left( \int\limits_{\lambda _{2}}^{\lambda _{3}}F\left(
\lambda
\right) d\lambda \right)
\end{equation}

\begin{equation}
W2=-2.5\lg \left( \int\limits_{\lambda _{1}}^{\lambda _{4}}F\left(
\lambda
\right) d\lambda \right)
\end{equation}

\begin{equation}
C=-2.5\lg \left( \frac{\lambda _{4}-\lambda _{1}}{\lambda _{3}-\lambda
_{2}}%
\right) ,
\end{equation}

\noindent
$F\left( \lambda \right) $ - is the spectral energy distribution,
$\lambda_{1},\lambda_{4}$ are the bluest and reddest wavelengths
of $W2$ respectively, while $\lambda_{2},\lambda_{3}$ are wavelength limits
for $W1$. We designed the filter $W2$ to be wide enough, while allowing to
avoid $H\beta$ at its blue side and the atmospheric lines at its red side.
The filter $W1$ has been designed to cover as large $v_{rad}$ interval
as possible and to allow a plausible photometric dynamical range for $W$.
However, we neglected the influence of the emission line O{\sc III} at
$500.7~nm$ because our main targets are early type galaxies. A redefinition
of the index $W$ for late type galaxies is easy and works well but then only
a small range of radial velocities can be covered using one set of filters.
These obviously contradicting conditions restricted the number of possible
solutions for the definition of the final system.

Figure 1 shows the main properties of the new system. Panels (a,b,c) demonstrate
a strong relation of the photometric index $W$ with the spectroscopic indices
$Mg_{1}$ and $Mg_{2}$ (as defined in Worthey et al. 1994). All these indicators
are (and should be) measured employing only the flux-calibrated spectra. We used
the stellar sample of Jones (1997) ($\sim600$ stars), the spectra of globular
clusters (Covino et al. 1995) and galaxies (Gorgas et al. 1997), the galaxy
templates and individual spectra (types $E-Sb$) from Kinney et al. (1996), and
the single-age single-metallicity models (SSPs) (Vazdekis 1999). We find
a linear relation between $W$ and $Mg_{2}$. However, $Mg_{1}$ (shown in Fig.~1)
and $Mg_{b}$ (Worthey et al. 1994) exhibit a non linear behavior with $W$, and
therefore cannot be simply applied for a calibration of $W$ because a nonlinear
transformation to the {\em standard} system would be required.

\begin{figure}
\caption{$Mg_{1}$ and $Mg_{2}$ vs. $W$ for a) stars, b) galaxies \& globular clusters,
c) galaxy models.}
\end{figure}

\section {Properties of the System}

Figure 2 shows a behavior of the index $W$ for the SSPs when definition of
the filters $W1$ and $W2$ is varied within a range of an achievable manufacturing
accuracy. Panel (a) demonstrates the changes of $W$ when $\delta\lambda$
and $\lambda_{0}$ characterizing $W2$ are varied, while keeping constant the
parameters of the filter $W1$. Panel (b) shows the changes of $W$ when parameters
of the filter $W1$ are varied, while for $W2$ they remain standard. From Figure 2
we learn that the linearity is independent on the precise definition of the filters,
and the required corrections for the analyzed cases are rather small. Therefore the
proposed system is quite robust against small changes of the main characteristics
of the filters.

\begin{figure}
\caption{The index $W$ vs. $Mg_{2}$ for different definitions of the filters. The
parameters of the filters $W2$ \& $W1$ are changed in the panels a) and b) respectively.}
\end{figure}

Since galaxies have different radial velocities we should study the stability
of the system response against the spectrum shift and constrain the range of
$v_{rad}$ where the {\em standard} system is applicable.
Figure 3 shows $W$ versus $Mg_{2}$ for the spectra of SSPs redshifted assuming
various $v_{rad}$. We do not see significant deviations except for
$v_{rad}=3000~km/s$. Once again only small linear correction factors should be
applied to $W$ allowing an accurate transformation to the {\em standard} system.
We see that the strongest correlation is for $v_{rad}\sim 1000~km/s$, the value
for which the system was optimized.

\begin{figure}
\caption{The index $W$ vs. $Mg_{2}$ for different recessional velocities applied to
the galaxy models. Filter set is standard.}
\end{figure}

The observing run must include observation of several standard stars (or, even
better, standard fields) with $Mg_{2}$ values determined accurately from
flux-calibrated spectra. The correction for the atmospheric extinction and all
data reduction steps dealing with CCD camera's and filter's distortions should
be carefully performed. The standard stars, SSP models and availability of
galaxy radial velocity determination will warrant an accurate transformation
to the {\em standard} system at $v_{rad}=0~km/s$. The dependence of
the transmission curves of real filters on temperature and focal ratio of the
telescope should also be taken into account.

\section{Advantages and Disadvantages of the New System}

The main advantages of the newly introduced photometric system are:
a strong linear relation with respect to the spectroscopic index $Mg_{2}$;
large range of $v_{rad}$ covered by one set of filters; efficiency to
reach high $S/N$ within short exposure times. A strong linear relation
of $W$ with $Mg_{2}$ ensures an accurate transformation of the observations
to the {\em standard} system, independent on the galaxy radial velocities up
to $v_{rad}\sim 2000~km/s$. The system could be easily reproduced at any
observatory if adequate calibration stars are observed. The stability of
the system within a wide range of $v_{rad}$ is also very important for
observations of the galaxies possessing a large internal radial velocity
differences. In addition, we see a great advantage in having a large field
of view which is limited only by the CCD and filter dimensions, and the
image scale in the focal plane of telescope.

Unfortunately, narrow photometric dynamical range of $W$ requires very accurate
correction for the atmospheric extinction, high photometric precision of the data,
and precise transformation to the {\em standard} system. Additional limitation of
the system in present definition is its applicability for the analysis of
early type ($E-Sb$) galaxies only.

\begin{acknowledgements}
Many thanks to A. Ku\v{c}inskas for helping us to clarify some points under
discussion.
\end{acknowledgements}

\end{article}

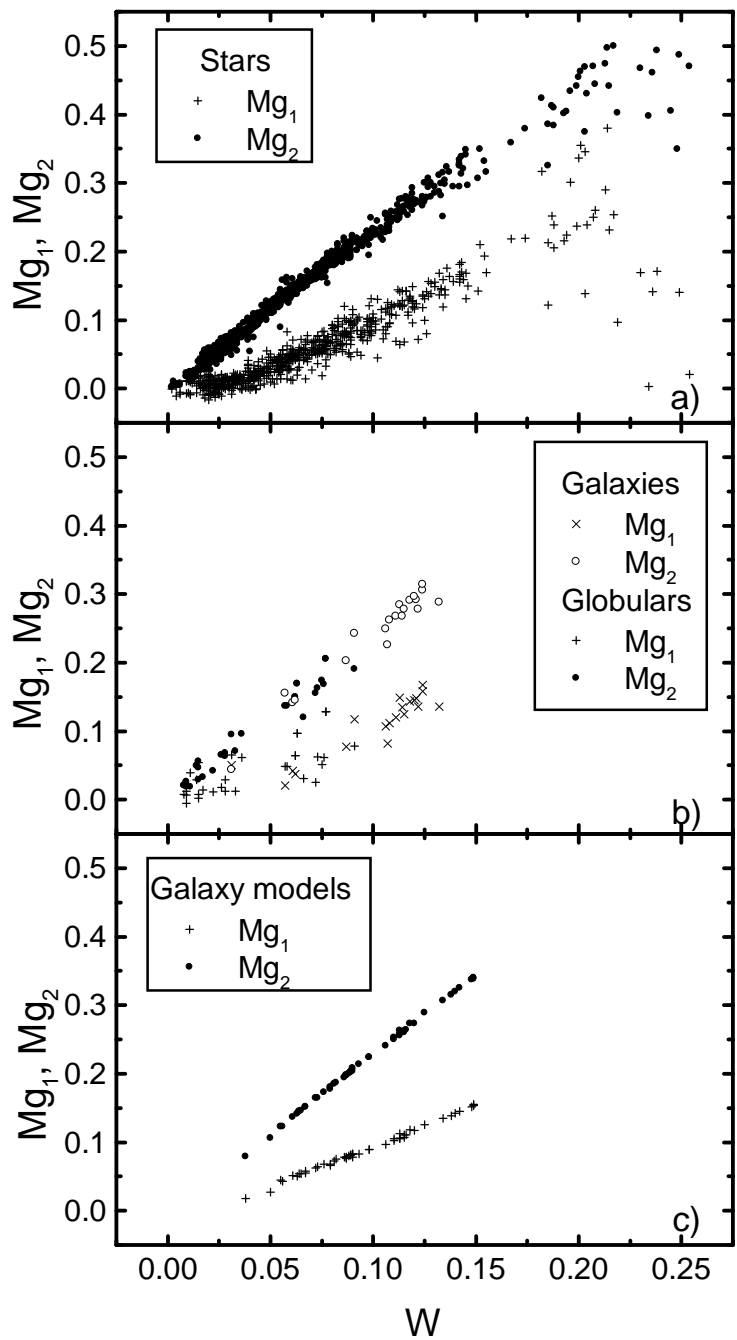
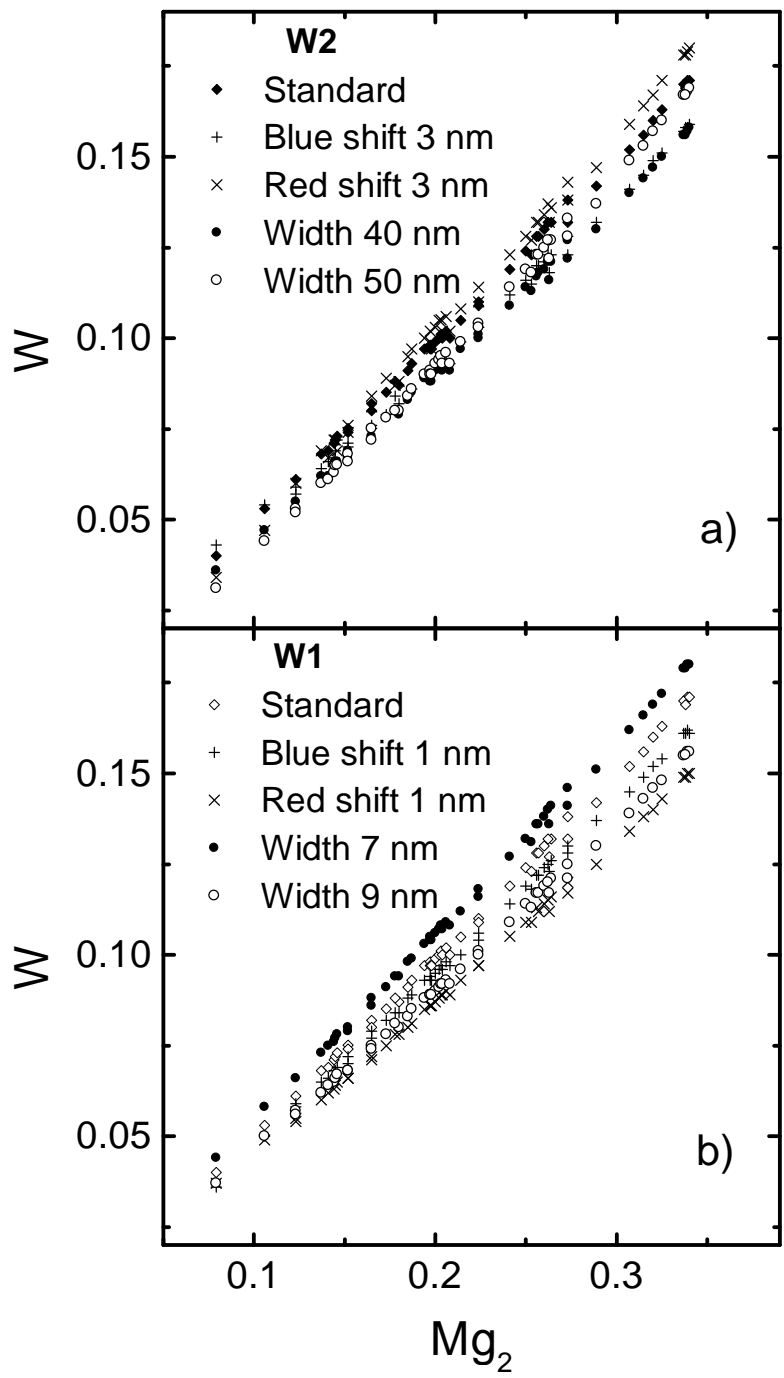
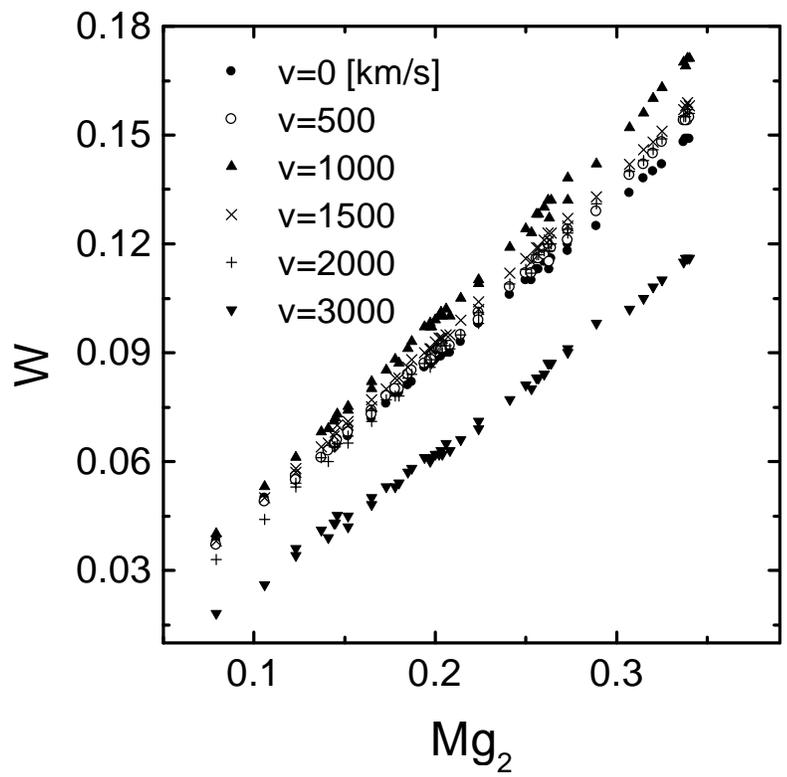
\end{document}